\documentclass[doublecol]{epl2}
\usepackage{graphicx}% Include figure files
\usepackage{dcolumn}% Align table columns on decimal point
\usepackage{bm}% bold math
\usepackage{colortbl}
\usepackage{amsmath, amsfonts, amssymb}

\title{Electronic correlations in double ionization of atoms in pump-probe experiments}

\author{S. Bauch  \and K. Balzer \and M. Bonitz }
\shortauthor{S. Bauch \etal}
\institute{

Institut f\"ur Theoretische Physik und Astrophysik\\
Christian-Albrechts-Universit\"at zu Kiel, D-24098 Kiel, Germany
}%

\date{\today}% It is always \today, today,

\abstract{
The ionization dynamics of a two-electron atom in an attosecond XUV-infrared pump-probe experiment is simulated 
 by solving the time-dependent two-electron Schr\"odinger equation. A dramatic change of the double ionization (DI) yield with variation of the pump-probe delay is reported and the governing role of electron-electron correlations is shown. 
The results allow for a direct control of the DI yield and of the relative strength of double and single ionization.}

\begin{document}

\maketitle

\section{Introduction}
With the emergence of coherent, ultrashort laser pulses in the extreme ultraviolet (XUV) regime
the time-resolved investigation of ultrafast  processes inside atoms became experimentally within reach, often accounted for as `attophysics' \cite{krausz2009}. 
Typically, a higher-harmonics generated (HHG) sub-femtosecond XUV pump pulse triggers the electronic motion which is then probed by an infrared (IR) or optical pulse with adjustable delay.
Owing to the weak intensity of the XUV pulse, usually strong probing pulses are needed, which are nowadays routinely controllable and reproducible \cite{brabec2000}. 
These new techniques allow for the time-resolved exploration of fundamental sub-femtosecond dynamics in the electronic properties of atoms \cite{drescher2002,uiberacker2007,goulielmakis2004}, 
molecules \cite{baker2006} and atoms on surfaces \cite{bauer2005}.

Often the complicated treatment of
multi-electron effects is simplified utilizing a single-active electron (SAE) description, 
together with the sudden approximations for the XUV pulse. While this seems to yield satisfactory agreement
with a certain class of recent experiments \cite{kazansky, kazansky2}, the validity range of the SAE remains open.
As we will show, in many pump-probe scenarios (in particular, with intense probe pulses) two-electron effects are crucial and a full description of both, electron-electron (e-e) correlations and the two laser pulses, is necessary.

In this Letter we address two important multi-electron processes in atoms: (i) the laser induced shake-up effect during XUV photon absorption and (ii) the rescattering mechanism.
In (i) the XUV-photon removes one electron, causing a rapid change of the binding potential of the remaining electron(s). During their rearrangement, the XUV-photon energy is shared
between the outgoing and the remaining electron(s) which is a consequence of e-e correlations. 
On the other hand, (ii) is a continuum effect and plays an important role for double ionization (DI) in the strong field regime. This mechanism has been investigated in great detail for helium, e.g. \cite{walker1994},
 and the importance of e-e correlations as well as the responsible mechanisms for non-sequential (NS) DI have been subject of many studies, where finally experiments based on the COLTRIMS
(cold-target recoil-ion momentum spectroscopy\cite{doerner2000,ullrich2003}) technique favoured rescattering with impact ionization \cite{weber2000, weber2000b, moshammer2000}.  
However, still today not every facet of this process is understood \cite{mauger2010}. Most importantly, the time dependence of nonsequential DI and, in particular, its manifestation in a time-resolved pump-probe experiment, 
have remained unexplored. In this Letter, we study these questions. We demonstrate that mechanisms (i) and (ii) are of key importance for DI and that they are intimately coupled via electron energy transfer.
 Moreover, double ionization can be controlled to a high degree via mechanism (ii)\footnote{The controllability of mechanism (ii) through carrier-envelope phase modulation of a single pulse has been demonstrated in 
\cite{liu2004} and within a two-color few-cycle situation in \cite{zhou2010}. However, we focus on control by delay and intensity in an XUV-IR two-pulse experiment.}.

Finally, we predict, that time-resolved observation of such processes is experimentally within reach. 

%-------------------
\section{Method}
The simplest system where the above mentioned effect are expected to occur is a two electron atom.
 Let us consider two electrons in a binding potential $V_{\textup{bi}}$ and a time-dependent perturbing field $V_{\textup{ex}}$. Their motion is described by the 
two-particle time-dependent Schr\"odinger equation (TDSE), which reads 
\begin {eqnarray}
 &&i\partial_t \Psi(\boldsymbol{r}_1,\boldsymbol{r}_2,t)=\left \lbrace -\frac{1}{2} \left (\partial_{\boldsymbol{r}_1}^2+\partial_{\boldsymbol{r}_2}^2 \right ) + \right. \label{eq:tdse} \\
  && \left. \sum_{i=1}^2\left [V_{\textup{bi}}(\boldsymbol{r}_i)+V_{\textup{ex}}(\boldsymbol{r}_i,t) \right ]+w(|\boldsymbol{r}_1-\boldsymbol{r}_2|)\right \rbrace \Psi(\boldsymbol{r}_1,\boldsymbol{r}_2,t) \; .\nonumber
\end {eqnarray}
All quantities are given in atomic units (a.u.) throughout this work ($\hbar=m_e=|e|=1/4\pi \epsilon_0 = 1$), unless stated explicitly. 
Equation~\eqref{eq:tdse} in its full dimensionality has been solved for various strong field situations, see e.g.~\cite{taylor2005, parker2006} and references therein. 
However, for an adequate treatment of the excitations considered in this work, the solution of the full two-electron problem is, yet, not manageable.
Therefore, it is reasonable to consider a one-dimensional model atom, which has been successfully applied to describe strong-field interaction with helium, e.g. \cite{lein2000,su1991, haan1994}. 
However, due to its long-range character the 1D Coulomb binding potential still introduces high complexity in the description of the two-electron continua giving rise to
a high computational effort (extremely large grids are required to simulate XUV-IR pump probe scenarios).

\begin{figure}
 \onefigure[width=0.48\textwidth]{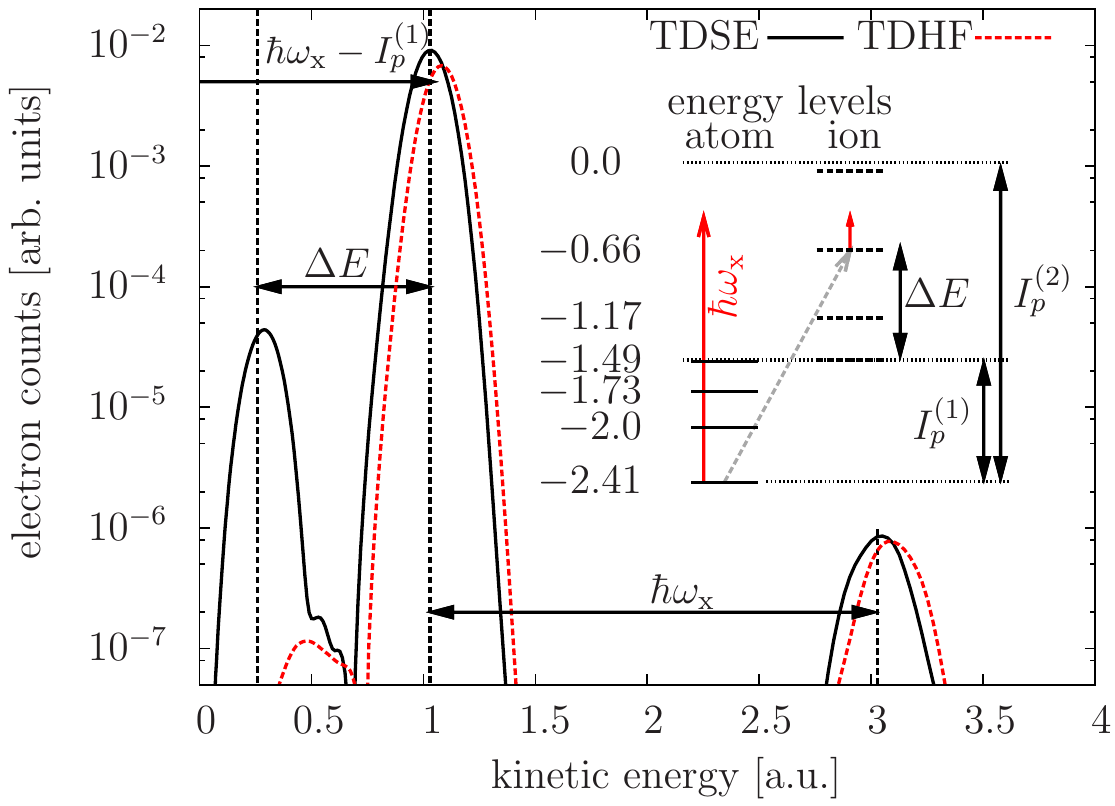}
\caption{(color online) Kinetic energy distribution of an XUV excited photoelectron from TDSE, (black) solid lines, and TDHF calculations, (red) dashed lines.
The inset depicts the (exact) two-electron and the corresponding single-electron (ion)  energy spectrum. The photon energy $\hbar \omega_{\textup{x}}$, schematically indicated by the (red) long arrow, 
is in between the first and second ionization threshold $I_p^{(1)}$ and $I_p^{(2)}$. The dominant shake-up process with energy transfer $\Delta E$ to the second excited ion state is indicated by the (gray) dashed arrow.
}
\label{fig:kineticenergies}
\end{figure}

In order to avoid these long-range binding effects, here, $V_{\textup{bi}}(x_i)$ is modeled by a 1D potential well of width $2a$ and depth $v_0$ where $V_{\textup{bi}}=0$
outside $[-a,a]$. We will show below (cf. Conclusions) that this model correctly reproduces key features of the two-color ionization dynamics of 1D helium. We use a depth of $v_0=1.6$ and a width of $2a=5.6$ which allows for a reasonable description of a helium-like atom, including
the ground state energy and a sufficient number of bound two-electron and single-electron (ion) eigenstates. 
A similar model has recently been used to compute the attosecond (as) electron pump-probe dynamics of surface atoms in solids \cite{krasovskii, krasovskii2}. 
%We note that this model, if parameters are chosen carefully, describes real atoms, besides the long-range binding, which is not essential for our investigations presented in this letter.
%
In the spirit of the one-dimensional helium model, the electrons interact via a softened Coulomb potential, $w(x_1-x_2)=1/\sqrt{(x_1-x_2)^2+1}$.
With that, the ionization potentials for single ionization (SI) and DI evaluate to $I_p^{(1)}=0.92$ and $I_p^{(2)}=2.41$, cf. inset in Fig.~\ref{fig:kineticenergies}.

\begin{figure}
 \onefigure[width=0.48\textwidth]{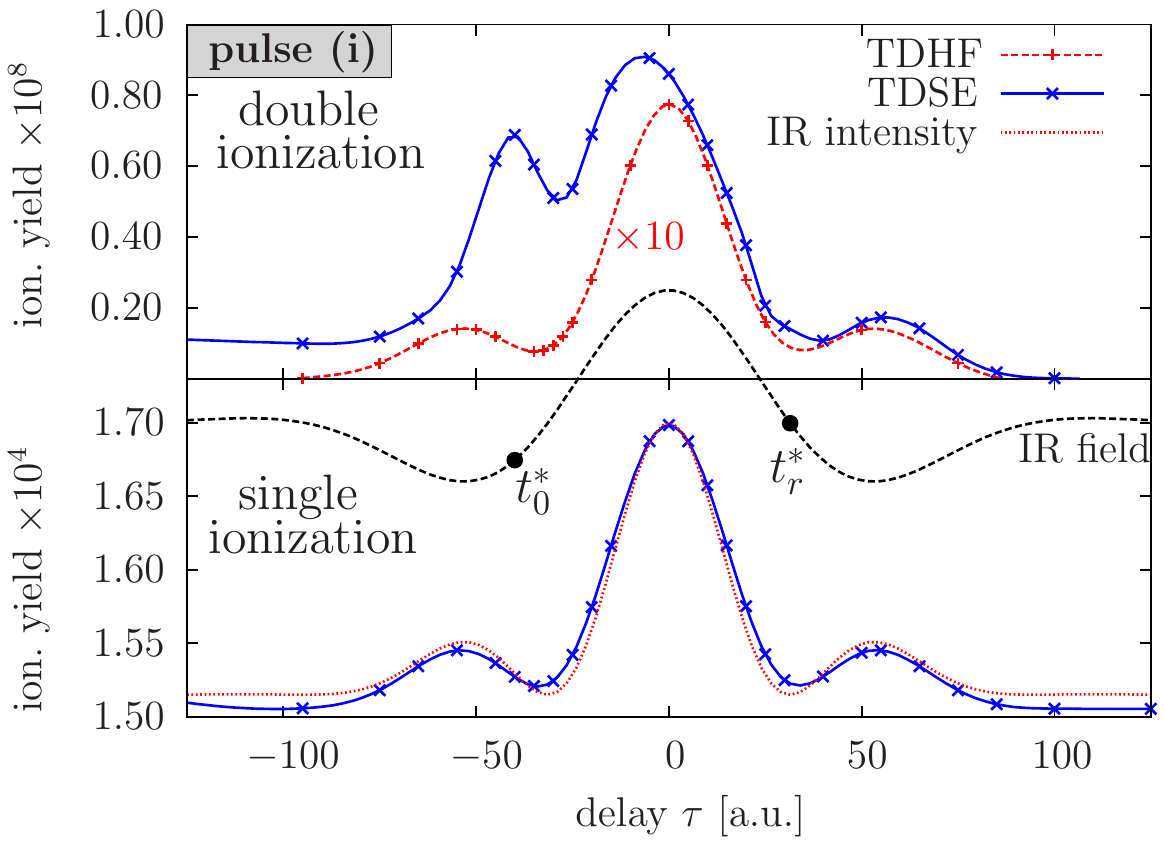}
 \caption{(color online) Single and double ionization yield versus delay time $\tau$ obtained from solutions of the TDSE and within TDHF, respectively, for an IR probing laser intensity of $I=8.8\cdot 10^{13}\;
 \un{W/ cm^2}$, [pulse (i)]. The DI graphs are normalized to zero for large delays (sequential DI). Furthermore, the TDHF curve is scaled by a factor $10$. The temporal laser intensity profile, $\propto |E(t)|^2$, is sketched by the (red) dotted line in the lower figure. The maximum impact energy is reached by an electron created at $t_0^*$ and rescattering at $t_r^*$.}
\label{fig:ionizationpulse1}
\end{figure}

The two-color laser field is treated within the dipole approximation with Gaussian envelopes 
%with fixed carrier-envelope phase is used, which gives
%
\begin {eqnarray} 
     V_{\textup{ex}}(x_i,t) &=& -x_i \left \lbrace E^0_{\textup{x}} \exp \left( -\frac{(t-\tau)^2}{2 \sigma^2_{\textup{x}}}\right) \cos [\omega_{\textup{x}} (t-\tau)] \right.  \nonumber \\
                     &+& \left. E^0_{\textup{ir}} \exp\left( -\frac{t^2}{2 \sigma^2_{\textup{ir}}} \right ) \cos (\omega_{\textup{ir}} t)  \right \rbrace \;.
\nonumber
%\label{eq:irxuvpulse}
\end {eqnarray}
Throughout, we use a $240\un{as}$ ($\sigma_x=10$) pump pulse with a photon energy of  $I_p^{(1)}< \hbar\omega_{\textup{x}}=1.99$ ($54 \un{eV}$) $<I_p^{(2)}$ and intensity of $8.8 \cdot 10^{13} \;\un {W/cm^2}$. These parameters are chosen such that significant SI is observed but multi-photon (MP) absorption is suppressed. 
The probing pulse is given by a strong $900\un{nm}$ few-cycle laser pulse, cf. Fig.~\ref{fig:ionizationpulse1}. The delay time $\tau$ of the XUV pulse is defined with respect to the maximum IR intensity and
varied in temporal steps of the order of $\sigma_x$.

We solve Eq.~\eqref{eq:tdse} in coordinate representation within a grid-based finite-difference method \cite{hoboken2010} for a series of delay times $\tau$.
Due to high kinetic energies occurring in our simulations, as a result of XUV photon absorption and subsequent IR acceleration, large spatial grids are needed to avoid reflections at the boundaries. Typically, we 
use a box size of at least $-500 < x_i < 500$  with a minimum of $8192$ grid points for each particle. 
To achieve proper time resolution ($\tau$-dependence) we have performed about $50$ runs for two different IR pulses, see below.
As initial state, we choose the spin-singlet ground state which possesses a symmetric spatial wave function, 
$\Psi(x_1,x_2,t)=\Psi(x_2,x_1,t)$. 

 The observables of interest are the SI and DI yield defined as the probability to find, respectively,
one and two electrons outside a certain (sufficiently large) distance from the binding potential, see e.~g. \cite{dahlen2001, camiolo2009}. 
 Due to the absence of long-range binding and delocalized states in our model the present procedure is particularly well suited to compute SI and DI and to discriminate between both.
Kinetic energy spectra of photoelectrons coming from singly-ionized systems are obtained by projecting the final two-electron wave function onto plane waves (single-particle states), integrating
out the bound part of the two-body wave function.

\section{Results}
 Let us first consider an IR intensity of $8.8 \cdot 10^{13} \; \un{W/cm^2}$ [pulse (i)]. The corresponding SI and DI yields for different delays $\tau$ are shown in Fig.~\ref{fig:ionizationpulse1}. 
We first note that SI directly follows the instantaneous intensity of the IR pulse at time of XUV excitation. This is not surprising, since the additional electrical field from the IR pulse increases the
resulting total intensity of the combined IR and XUV pulse and the XUV photon energy $\hbar \omega_x$ is sufficient to ionize the system independently of the IR field strength.
The DI, however, shows a completely different $\tau$-dependence. While for $\tau>0$ it follows the IR intensity, 
for negative delays it strongly departs from the latter. In contrast to SI the curve is not symmetric with respect to $\tau=0$, and even shows a significant maximum at the leading edge
of the IR pulse, around $\tau \approx -40$. 
Since $\hbar \omega_x < I_p^{(2)}$, the DI is truly caused by non-linear processes,
either by (at least) two XUV photon absorption or through a combined action of both pulses. Since the two-photon absorption is a sequential effect, the DI should in that case follow the form of the SI yield,
which is obviously not the case, cf. Fig.~\ref{fig:ionizationpulse1}. Further, as the IR intensity is too low for DI through strong field tunneling or MP ionization, only the combination of both pulses can be responsible for DI.

\section{XUV induced shake up}
 We first consider the case of XUV-only excitation. 
Two different methods are employed: the full solution of the two-particle TDSE, Eq.~(\ref{eq:tdse}), and the solution within the time-dependent Hartree-Fock
 (TDHF) approximation \cite{balzerfedvr} neglecting, by definition, all e-e correlation effects. For recent comparisons of time-dependent mean-field calculations with the exact treatment of the
 TDSE in the context of strong-field interactions addressing limitations and improvements, see e.g. \cite{ruggenthaler2009,balzer2010,dewijn2008,wilken2007}.

The kinetic energy spectrum of photoelectrons coming from singly-ionized systems  is shown in Fig.~\ref{fig:kineticenergies}. Two distinct peaks separated by the photon energy $\hbar \omega_x$
 can be identified in both spectra, TDHF and TDSE, at the expected positions of $\hbar \omega_x -I_p^{(1)} \approx 1$, for absorption of a single photon,
and $2 \hbar \omega_x -I_p^{(1)} \approx 3$, for two-photon absorption, respectively. Ground state e-e correlations become apparent in a small general shift between TDHF and TDSE spectra ($E_0^{\textup{TDSE}}=-2.41$ and $E_0^{\textup{TDHF}}=-2.35$).
A remarkable feature is the presence of a third pronounced peak at an energy of  $E_{\textup {kin}}\approx 0.28$ in the TDSE calculations, corresponding to slow photoelectrons.
The missing kinetic energy, $\Delta E=E_{\textup{shakeup}}=0.83$, can be directly associated with the  energy difference from the ion ground state to the second excited state. Therefore, and from the fact
that this contribution is absent in TDHF, cf. Fig.~\ref{fig:kineticenergies}, we conclude that this peak is 
direct evidence of the population of shake up states by the remaining
electron in the ion\footnote{In fact, a careful inspection of the DI in Fig.~\ref{fig:ionizationpulse1} reveals, that this shake-up state can be depopulated via IR probing, noticeable
from the fact that the yield for $\tau<-100$ is larger (state depopulated by IR pulse) than for $\tau>100$. This is in analogy to the experiment by Uiberacker \emph{et al.} \cite{uiberacker2007}
where shake-up state population is probed in a similar way.}, see inset of Fig.~\ref{fig:kineticenergies}. \\

\section{Semiclassical rescattering model (SRM)} 
To shed more light into the underlying physics, we follow a classical analysis of the electron trajectories in the continuum \cite{corkum1993}. 
In many cases, a (semi-) classical picture turns out to describe a variety of strong-field processes. We mention here the discussion of 
above-threshold ionization\cite{goreslavski2004} and NSDI \cite{goreslavski2001,popruzhenko2000}
 incorporating Coulomb effects \cite{faria2004} and using few-cycle pulses \cite{liu2004b,faria2004b}. In this Letter we restrict ourselves
 to an adapted, simple version of the common three-step model:
Consider an electron ``created'' at time $t_0$ in the continuum with momentum $p_0$ [$t_0$ coincides with the delay time $\tau$]
which follows from energy conservation:
 $p_0 = \pm [2(\hbar\omega_{\textup{x}}-I_p^{(1)}-E_{\textup{shakeup}})]^{1/2}$. 
Here we took into account a possible kinetic energy loss due to shake up of
the second electron. 

Within this model,
the electron which was excited by the XUV pulse with momentum $p_0$, acquires 
in the IR field with vector potential $A(t)$ the momentum
\begin {equation}
 p(t,t_0) = p_0 + \frac{1}{c} A(t)= p_0 - \int_{t_0}^t \upd \overline{t} \; E(\overline{t}) \,, 
 \label{eq:classical_momentum}
\end {equation}
leading to the classical trajectory [$x(t_0)=0$],
\begin{equation}
 x(t,t_0)= p_0 (t-t_0) - \int_{t_0}^{t} \upd \overline{t} \int_{t_0}^{\overline{t}} \upd \bar{\bar{t}} \,E(\bar{\bar{t}}) \;.
\label{eq:trajectories}
\end{equation}
The condition $x(t_r,t_0)=0$, for $t_r>t_0$, gives the time of rescattering $t_r$ and
%the impact momentum $\boldsymbol{p}(t_r, t_0)$ and 
the impact energy at $t_r$: $E_{\textup{SRM}}(t_r; t_0)=p(t_r, t_0)^2/2$.
\begin{figure}
 \onefigure[width=0.47\textwidth]{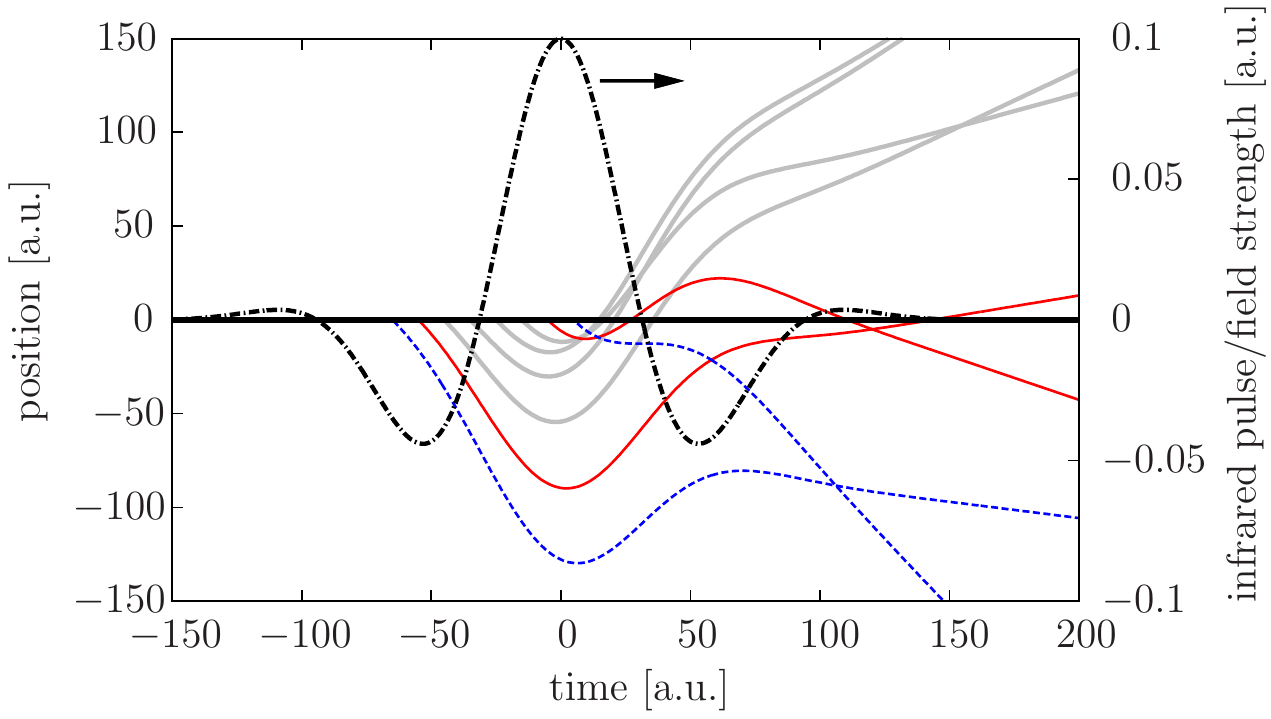}
\caption{(color online) Classical trajectories of an electron in the continuum created at time $t_0$ during the IR probing field (ii), cf. Eq.~\eqref{eq:trajectories}. Dashed (blue) lines
correspond to escaping electrons, (red and grey) solid lines to returning electrons. The probing IR field is indicated by the (black) dashed-dotted line.}
\label{fig:classical_trajectories}
\end{figure}
 
Solutions of Eq.~\eqref{eq:trajectories} for different times $t_0$, initial momentum $p_0=\sqrt{2 E_{\textup{kin}}}$ extracted from the kinetic energy distribution in Fig.~\ref{fig:kineticenergies}
and an IR intensity of $3.5 \cdot 10^{14} \; \un{W/cm^2}$ are shown in Fig.~\ref{fig:classical_trajectories}.
Trajectories for returning electrons (solid lines) are allowed during the ascending cycle of the IR pulse, whereas electrons created at different times (dashed lines) escape.

\begin{figure}
 \onefigure[width=0.47\textwidth]{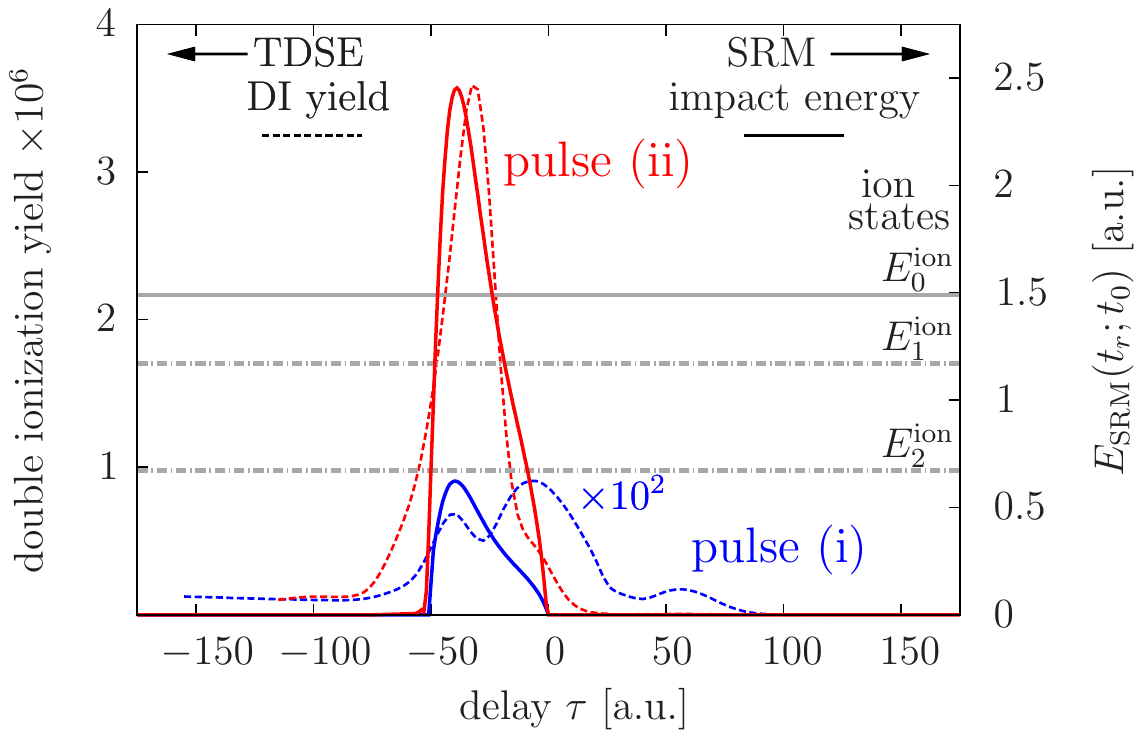}
\caption{(color online) Impact energy (full lines) of the XUV-excited electron created at $t_0 \equiv \tau$
for the IR intensities $I=8.8\cdot 10^{13}\; \un{W/cm^2}$  [pulse (i)] and $I=3.5 \cdot 10^{14}\; \un{W/cm^2}$ [pulse (ii)].  For comparison,
 the double ionization yield from TDSE simulations (cf. Figs.~\ref{fig:ionizationpulse1} and ~\ref{fig:ionizationpulse2}) is shown by dashed lines. 
Horizontal gray lines mark the possible binding energies of the remaining electron (eigenstates of ion), i.~e.~the minimal kinetic energy of the returning electron necessary for impact double ionization.}
\label{fig:returning_energy}
\end{figure}

The analysis of Eq.~(\ref{eq:trajectories}) reveals that, for  IR pulse (i), 
only electrons with $p_0>0$ and initial kinetic energy $p_0^2/2 \lesssim 0.4$ [cf. Fig.~\ref{fig:kineticenergies}] can be driven back to the ion by the IR field thereby obtaining a
kinetic energy $E_{\textup{SRM}}(t_r)$ sufficient for impact ionization. The energy $E_{\textup{SRM}}$ is shown in Fig.~\ref{fig:returning_energy} versus
$t_0$ and exhibits a maximum around $\tau=t_0^*=-40$. 
In fact, an electron ``created'' around $t_0^*$ will be accelerated until $t_r^*$, i.e. for the entire half  cycle of positive IR field strength, 
cf. Eq.~(\ref{eq:classical_momentum}) and Fig.~\ref{fig:ionizationpulse1}.
Interestingly, the maximum position of $E_{\textup{SRM}}(\tau)$ is very close to that of the maximum of the DI yield observed in the TDSE simulations, cf.  Fig.~\ref{fig:returning_energy}, 
clearly supporting the rescattering mechanism.
Figure~\ref{fig:returning_energy} also shows that the value of $E_{\textup{SRM}}$ is sufficient for impact ionization from the shake up state\footnote{The peak of $E_{\textup{SRM}}$ refers to electrons with an initial energy $p_0^2/2 \approx 0.28$ 
(vertical dashed line in Fig.~\ref{fig:kineticenergies}), whereas electrons with a smaller initial energy will return with a larger value $E_{\textup{SRM}}$ exceeding $E_2^{\rm ion}$.}, $E_2^{\rm ion}$.

Note that the dominant fraction of electrons (with energies around $1$, cf. Fig.~\ref{fig:kineticenergies}) escapes without rescattering. Hence, the energy loss due to shake up of the second electron is crucial for impact ionization.
%
%Therefrom, we conclude that the main DI peak around $\tau=-40$ originates from the rescattering of one electron impact ionizing the second one.
To further verify the rescattering mechanism we have performed a series of TDHF calculations for the whole range of delays $\tau$.
Within TDHF both, SI {\em and DI},  follow the IR intensity profile, cf. Fig.~\ref{fig:ionizationpulse1}, in contrast to full TDSE solutions.
Thus, e-e correlation effects are responsible for the observed $\tau$-dependence of the DI yield.

\section{Controlling double ionization}
\begin{figure}
 \onefigure[width=0.48\textwidth]{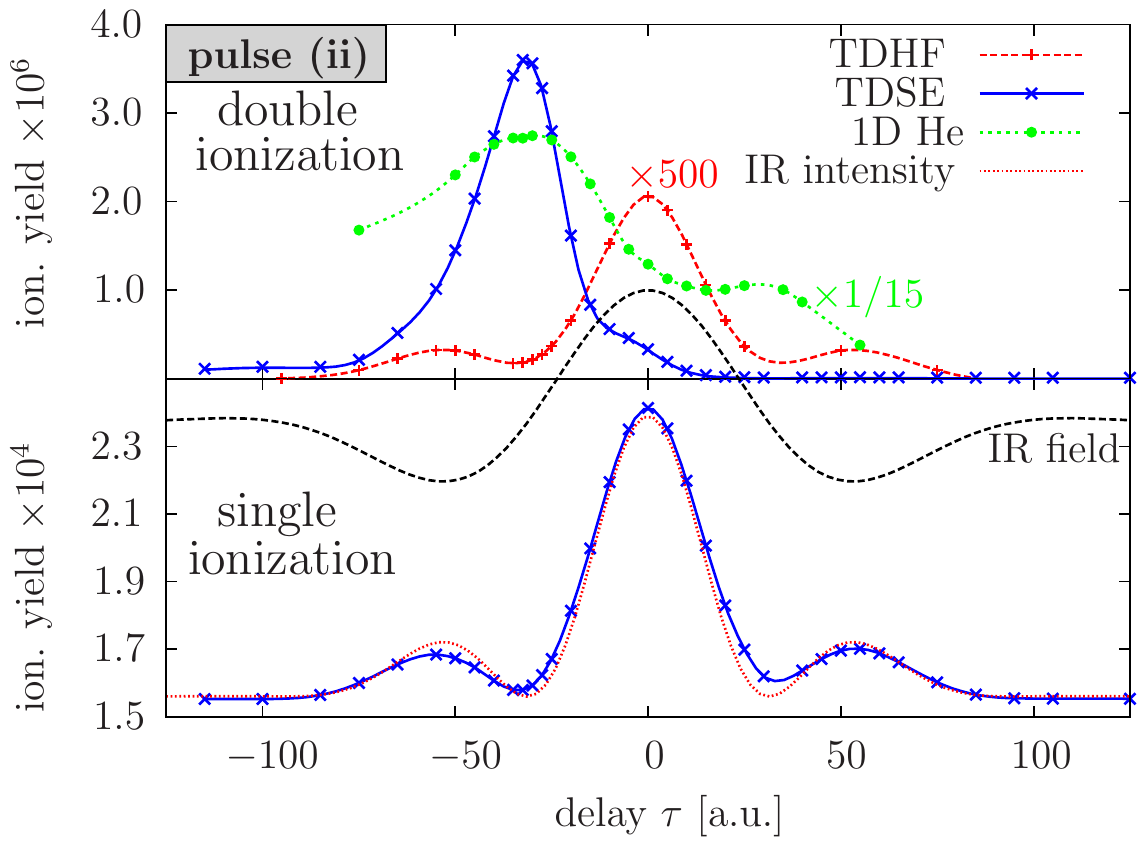}
 \caption{(color online) Same as Fig.~\ref{fig:ionizationpulse1}, but for an IR  intensity of $I=3.5\cdot 10^{14}\;\un{W/cm^2}$, 
pulse (ii). Additionally, the results for a 1D helium model are shown for comparison in the upper figure.}
\label{fig:ionizationpulse2}
\end{figure}
Now the question arises whether it is possible to further increase the DI yield relative to SI.
From the above analysis it is obvious that DI could be enhanced if 
the dominant contribution of XUV photo-excited electrons, i.~e.~ electrons with $E_{\textup{kin}}(t_0)\sim 1$ in Fig.~\ref{fig:kineticenergies}, would be able to impact ionize the remaining electron.
In fact, this is easily achieved by increasing the IR intensity, as we demonstrate below 
for  $I=3.5\cdot 10^{14}\;\un{W/cm^2}$ [pulse (ii)].
As before, SI follows the IR intensity profile, and DI shows a peak around $\tau=-40$ which, however, 
is now dramatically enhanced, cf. Fig.~\ref{fig:ionizationpulse2}. 
With the present intensity increase by a factor of four, an enhancement of DI by more than two orders  of magnitude is achieved, whereas SI is not affected significantly. 
The comparison with the classical impact energy, cf. Fig.~\ref{fig:returning_energy}, shows again an excellent agreement between the maximum positions of $E_{\textup{SRM}}$ and the DI yield. Also, 
the peak height of $E_{\textup{SRM}}$ now by far exceeds even the ground state energy of the electron in the ion, thus now DI occurs from all ionic bound states.
The complete absence of this peak in TDHF calculations (which predict a DI yield which is three to four orders of magnitude too small), cf. Fig.~\ref{fig:ionizationpulse2}, 
proves again that  e-e correlations are the origin of this effect.

\section{Conclusions}
In conclusion, we have demonstrated that in XUV-IR pump-probe scenarios DI proceeds via a combination of electron shake-up and rescattering, at low IR intensity,
or via rescattering at high IR intensity. By properly choosing the delay between the XUV and IR pulse and intensity of the IR pulse the DI yield can be varied within two to three orders of magnitude, reaching up to several percent of the SI yield. A simple physical explanation has been 
given which is straightforwardly extended to other laser pulse shapes.
We verified that our predictions for a 1D model atom can be directly applied to small atoms. This is supported by a series of two-electron TDSE simulations for a 1D helium atom for the parameters used  in this letter.
The results are included in Fig.~\ref{fig:ionizationpulse2}. Obviously, our main observation, the non-monotonic delay dependence of DI, is correctly captured by the model atom. This indicates that the correlation-dominated
DI mechanism is present also in case of a Coulomb potential. On the other hand, there are quantitative differences (note the 11-fold increase of the peak height) which can be attributed to differences in the excitation spectra.
A more detailed analysis of XUV-IR excitation of 1D helium requires very large computational effort and is subject of ongoing work \cite{bbb2010}.
Since the laser parameters used in our calculations are typical for current experimental
conditions, we expect that time-resolved measurements of this process should be possible and allow to shed more light into one of the most intriguing manifestations of correlated electron dynamics in nature.

\acknowledgments
This work has been supported by the North-German super-computer alliance (HLRN) via grant shp0006 and, in part, by the Deutsche Forschungsgemeinschaft via SFB-TR 24.


\begin{thebibliography}{999}

 \bibitem{krausz2009} 
    \Name{Krausz F. \and Ivanov M.}
    \Review{Rev. Mod. Phys.}
    \Vol{81}
    \Page{163}
    \Year{2009}

 \bibitem{brabec2000}
    \Name{Brabec Th. \and Krausz F.}
    \Review{Rev. Mod. Phys.}
    \Vol{72}
    \Page{545}
    \Year{2000}

\bibitem{uiberacker2007} 
  \Name{Uiberacker M., Uphues Th., Schultze M., Verhoef A. J., Yakovlev V., Kling M. F., Rauschenberger J., Kabachnik N. M., Schr\"oder H., Lezius M., Kompa K. L., Muller H. G., Vrakking M. J. J., Hendel S., Kleineberg U., Heinzmann U., Drescher M. \and Krausz F.}
  \Review{Nature}
  \Vol{446}
  \Page{627}
  \Year{2007}

\bibitem{goulielmakis2004} 
   \Name{Goulielmakis E., Uiberacker M., Kienberger R., Baltuska A., Yakovlev V., Scrinzi A., Westerwalbesloh Th., Kleineberg U., Heinzmann U., Drescher M. \and Krausz F.}
   \Review{Science}
   \Vol{305}
   \Page{1267}
   \Year{2004}

\bibitem{drescher2002}
  \Name{Drescher M., Hentschel M., Kienberger R., Uiberacker M., Yakovlev V., Scrinzi A., Westerwalbesloh Th., Kleineberg U., Heinzmann U. \and Krausz F.}
  \Review{Nature}
  \Vol{419}
  \Page{803}
  \Year{2002}

 \bibitem{baker2006} 
   \Name{Baker S., Robinson J. S., Haworth C. A., Teng H., Smith R. A., Chirila C. C., Lein M., Tisch J. W. G., Marangos J. P.}
   \Review{Science}
   \Vol{312}
   \Page{424}
   \Year{2006}

\bibitem{bauer2005} 
   \Name{Bauer M.}
   \Review{J. Phys. D: Appl. Phys.}
   \Vol{38}
   \Page{R253}
   \Year{2005} 

\bibitem{kazansky} 
   \Name{Kazansky A. K. \and Kabachnik N. M.}
   \Review{J. Phys. B: At. Mol. Opt. Phys.}
   \Vol{40}
   \Page{2163}
   \Year{2007};

\bibitem{kazansky2}
   \Name{Kazansky A. K. \and Kabachnik N. M.}
   \Review{J. Phys. B: At. Mol. Opt. Phys}
   \Vol{41}
   \Page{135601}
   \Year{2008}

\bibitem{walker1994} 
  \Name{Walker B., Sheehy B., DiMauro L. F., Agostini P., Schafer K. J., \and Kulander K. C.}
   \Review{Phys. Rev. Lett.}
   \Vol{73}
   \Page{1227}
   \Year{1994}


\bibitem{doerner2000}
   \Name{D\"orner R., Mergel V., Jagutzki O., Spielberger L., Ullrich J., Moshammer R. \and Schmidt-B\"ocking H.}
   \Review{Phys. Rep.}
   \Vol{330}
   \Page{95}
   \Year{2000}  

\bibitem{ullrich2003}
   \Name{Ullrich J., Moshammer R., Dorn A., D\"orner R., Schmidt L. Ph. H. \and Schmidt-B\"ocking H.}
   \Review{Rep. Prog. Phys.}
   \Vol{66}
   \Page{1463}
   \Year{2003}

\bibitem{weber2000}
   \Name{Weber Th., Weckenbrock M., Staudte A., Spielberger L., Jagutzki O., Mergel V., Afaneh F., Urbasch G., Vollmer M., Giessen H. \and D\"orner R.}
   \Review{Phys. Rev. Lett.}
   \Vol{84}
   \Page{443}
   \Year{2000}  

\bibitem{weber2000b}
   \Name{Weber Th., Giessen H., Weckenbrock M., Urbasch G., Staudte A., Spielberger L., Jagutzki O., Mergel V., Vollmer M. \and D\"orner R.}
   \Review{Nature (London)}
   \Vol{405} 
   \Page{658}
   \Year{2000}

\bibitem{moshammer2000}
   \Name{Moshammer R., Feuerstein B., Schmitt W., Dorn A., Schr\"oter C. D., Ullrich J., Rottke H., Trump C., Wittmann M., Korn G., Hoffmann K. \and Sandner W.}
   \Review{Phys. Rev. Lett.}
   \Vol{84}
   \Page{447}
   \Year{2000}  

\bibitem{mauger2010}
   \Name{Mauger F., Chandre C. \and Uzer T.}
   \Review{Phys. Rev. Lett.}
   \Vol{104}
   \Page{043005}
   \Year{2010} 

 \bibitem{liu2004} 
   \Name{Liu X., Rottke H., Eremina E., Sandner W., Goulielmakis E., Keeffe K. O., Lezius M., Krausz F., Lindner F., Sch\"atzel M. G., Paulus G. G.  \and Walther H.}
   \Review{Phys. Rev. Lett.}
   \Vol{93}
   \Page{263001}
   \Year{2004}

 \bibitem{zhou2010}
   \Name{Zhou Y., Liao Q., Zhang Q., Hong W. \and Lu P.}
   \Review{Optics Express}
   \Vol{18} 
   \Page{632}
   \Year{2010}

\bibitem{taylor2005}
   \Name{Taylor K. T., Parker J. S., Dundas D., Meharg K. J., Doherty B. J. S., Murphy D. S. \and McCann J. F.}
   \Review{J. Electron. Spectrosc. Relat. Phenom.}
   \Vol{144--147}
   \Page{1191--1196}
   \Year{2005}

\bibitem{parker2006}
   \Name{Parker J. S., Doherty B. J. S., Taylor K. T., Schultz K. D., Blaga C. I. \and DiMauro L. F.}
   \Review{Phys. Rev. Lett.}
   \Vol{96}
   \Page{133001}
   \Year{2006}


\bibitem{lein2000}
   \Name{Lein M., Gross E. K. U. \and Engel V.}
   \Review {Phys. Rev. Lett.}
   \Vol{85}
   \Page{4707}
   \Year{2000}

\bibitem{haan1994} 
   \Name{Haan S. L., Grobe R. \and Eberly J. H.}
   \Review{Phys. Rev. A}
   \Vol{50}
   \Page{378}
   \Year{1994} 

\bibitem{su1991}
  \Name{Su Q. \and Eberly J. H.}
  \Review{Phys. Rev. A}
  \Vol{44}
  \Page{5997}
  \Year{1991}


\bibitem{krasovskii} 
   \Name{Krasovskii E. E.  \and Bonitz M.}
   \Review{Phys. Rev. Lett.}
   \Vol{99} 
   \Page{247601}
   \Year{2007} 

\bibitem{krasovskii2} 
   \Name{Krasovskii E. E. \and Bonitz M.}
   \Review{Phys. Rev. A}
   \Vol{80}
   \Page{053421}
   \Year{2009}


\bibitem{hoboken2010} 
   \Name{Bauch S., Balzer K., Ludwig P. \and Bonitz M.}
   \Editor{Bonitz M., Horing N. \and Ludwig P.}
   \Book{Introduction to Complex plasmas}
   \Publ{Springer, Berlin}
   \Year{2010}

 \bibitem{dahlen2001}
   \Name{Dahlen N. E. \and van Leeuwen R.}
   \Review{Phys. Rev. A}
   \Vol{64} 
   \Page{023405}
   \Year{2001}

\bibitem{camiolo2009}
   \Name{Camiolo G., Castiglia G., Corso P. P., Fiordilino E. \and Marangos J. P.}
   \Review{Phys. Rev. A}
   \Vol{79}
   \Page{063401}
   \Year{2009}


\bibitem{balzerfedvr} 
    \Name{Balzer K., Bauch S. \and Bonitz M.}
    \Review{Phys. Rev. A}
    \Vol{81}
    \Page{022510}
    \Year{2010}


\bibitem{ruggenthaler2009}
   \Name{Ruggenthaler M. \and Bauer D.}
   \Review{Phys. Rev. Lett.}
   \Vol{102}
   \Page{233001}
   \Year{2009}

\bibitem{dewijn2008}
   \Name {De Wijn A. S., Lein, M. \and K\"ummel S.}
   \Review{Euro. Phys. Lett.}
   \Vol{84}
   \Page{43001}
   \Year{2008}

\bibitem{wilken2007}
   \Name{Wilken F. \and Bauer D.}
   \Review{Phys. Rev. A}
   \Vol{76}
   \Page{023409}
   \Year{2007}

\bibitem{balzer2010}
   \Name{Balzer K., Bauch S. \and Bonitz M.}
   \Review{Phys. Rev. A}
   \Vol{}
   \Page{accepted for publication}
   \Year{2010}

\bibitem{corkum1993} 
   \Name{Corkum P. B.}
   \Review{Phys. Rev. Lett.}
   \Vol{71}
   \Page{1994}
   \Year{1993}

\bibitem{goreslavski2004}
   \Name{Goreslavski S. P., Paulus G. G., Popruzhenko S. V. \and Shvestsov-Shilovski N. I.}
   \Review{Phys. Rev. Lett.}
   \Vol{93}
   \Page{233002}
   \Year{2004}

\bibitem{goreslavski2001}
   \Name{Goreslavskii S. P., Popruzhenko S. V., Kopold R. \and Becker W.}
   \Review{Phys. Rev. A}
   \Vol{64}
   \Page{053402}
   \Year{2001}

\bibitem{popruzhenko2000}
   \Name{Popruzhenko S. V. \and Goreslavskii S. P.}
   \Review{J. Phys. B: At. Mol. Opt. Phys.}
   \Vol{34}
   \Page{L239}
   \Year{2000}


\bibitem{faria2004}
   \Name{Figueira de Morisson Faria C., Liu X. \and Becker W.}
   \Review{Phys. Rev. A}
   \Vol{69}
   \Page{021402(R)} 
   \Year{2004}

\bibitem{liu2004b}
   \Name{Liu X. \and Figueira de Morisson Faria C.}
   \Review{Phys. Rev. Lett.}
   \Vol{92}
   \Page{133006}
   \Year{2004}

\bibitem{faria2004b}
   \Name{Figueira de Morisson Faria C., Liu X. Sanpera A. \and Lewenstein M.}
   \Review{Phys. Rev. A}
   \Vol{70}
   \Page{043406}
   \Year{2004}

\bibitem{bbb2010}
   \Name{Bauch S., Balzer K. \and Bonitz M.}
   \Review{to be published}
   \Year{2010}




\end{thebibliography}
\end{document}